\documentclass{aa}
\usepackage{epsfig,times}
\makeatletter
\def\@cite#1#2{(#1\if@tempswa , #2\fi)}
\def\@citex[#1]#2{\if@filesw\immediate\write\@auxout{\string\citation{#2}}\fi
  \def\@citea{}\@cite{\@for\@citeb:=#2\do
    {\@citea\def\@citea{;\penalty\@m\ }\@ifundefined
       {b@\@citeb}{{\bf ?}\@warning
       {Citation `\@citeb' on page \thepage \space undefined}}%
\hbox{\csname b@\@citeb\endcsname}}}{#1}}
\makeatother

\begin{document}

\title{X--ray broad-band study of the \\ symbiotic X--ray binary 4U 1954+31}

\author{N. Masetti\inst{1},
E. Rigon\inst{2},
E. Maiorano\inst{1,3},
G. Cusumano\inst{4},
E. Palazzi\inst{1},
M. Orlandini\inst{1},
L. Amati\inst{1}
and F. Frontera\inst{1,3}}

\institute{
INAF -- Istituto di Astrofisica Spaziale e Fisica Cosmica di Bologna, 
Via Gobetti 101, I-40129 Bologna, Italy (formerly IASF/CNR, Bologna)
\and
Dipartimento di Fisica, Universit\`a di Bologna, Via Irnerio 48,
I-40126 Bologna, Italy
\and
Dipartimento di Fisica, Universit\`a di Ferrara, via Saragat 1, I-44100
Ferrara, Italy
\and
INAF -- Istituto di Astrofisica Spaziale e Fisica Cosmica di Palermo, 
Via Ugo La Malfa 153, I-90146 Palermo, Italy (formerly IASF/CNR, Palermo)
}

\titlerunning{X--ray study of the symbiotic X--ray binary 4U 1954+31}
\authorrunning{Masetti et al.}

\offprints{N. Masetti, {\tt masetti@iasfbo.inaf.it}}

\date{Received 6 October 2006; Accepted 14 November 2006}

\abstract{We present results of several pointed X--ray broad band observations of 
the `symbiotic X--ray binary' 4U 1954+31 performed with the satellites 
{\it BeppoSAX}, {\it EXOSAT}, {\it ROSAT}, {\it RXTE} and {\it Swift}
between October 1983 and April 2006. We also studied the {\it RXTE} ASM 
data over a period of more than 10 years, from January 1996 to October 2006. 
Light curves of all pointed observations show an erratic behaviour with 
sudden increases in the source emission on timescales variable from 
hundreds to thousands of seconds. There are no indications of changes in 
the source spectral hardness, with the possible exception of the {\it RXTE} 
pointed observation.
Timing analysis does not reveal the presence of coherent pulsations or 
periodicities either in the pointed observations in the range from 2 ms to 
2000 s or in the long-term {\it RXTE} ASM light curve on timescales from 
days to years. The 0.2--150 keV spectrum, obtained with {\it BeppoSAX}, is 
the widest for this source available to date in terms of spectral coverage 
and is well described by a model consisting of a lower-energy thermal 
component (hot diffuse gas) plus a higher-energy (Comptonization) 
emission, with the latter modified by a partially-covering cold absorber 
plus a warm (ionized) absorber. A blackbody modelization of our 
{\it BeppoSAX} low-energy data is ruled out.
The presence of a complex absorber local to the source is also supported 
by the 0.1--2 keV {\it ROSAT} spectrum. {\it RXTE}, {\it EXOSAT} and 
{\it Swift} X--ray spectroscopy is consistent with the above results, but 
indicates variations in the density and the ionization of the local 
absorber. A 6.5 keV iron emission line is possibly detected in the 
{\it BeppoSAX} and {\it RXTE} spectra.
All this information suggests that the scenario that better 
describes 4U 1954+31 consists of a binary system in which a neutron star 
orbits in a highly inhomogeneus medium, accreting matter from a stellar 
wind coming from its optical companion, an M-type giant star.
}

\maketitle

\section{Introduction}

\begin{table*}
\caption[]{Log of the X--ray pointings on 4U 1954+31 analyzed in
this paper.}
\begin{center}
\begin{tabular}{llccrl}
\noalign{\smallskip}
\hline
\hline
\noalign{\smallskip}
\multicolumn{1}{c}{Satellite} & \multicolumn{1}{c}{Obs. start} &
Obs. start & \multicolumn{1}{c}{Exposure} & 
\multicolumn{2}{l}{On-source time (ks)} \\
 & \multicolumn{1}{c}{date} & time (UT) & \multicolumn{1}{c}{(ks)} & & \\
\noalign{\smallskip}
\hline
\noalign{\smallskip}
{\it EXOSAT}   & 22 Oct 1983 & 21:50:00 &   15.2 & 14.7 & (GSPC)\\
               &             & 21:53:04 &   14.8 & 13.4 & (ME)$^*$\\
               & 02 Jun 1985 & 06:54:08 &   16.3 & 14.8 & (GSPC)\\
               &             & 06:57:20 &    5.3 &  5.2 & (ME)\\
               &             & 08:37:36 &   10.2 &  8.9 & (ME)\\
\noalign{\smallskip}
{\it ROSAT}    & 04 May 1993 & 12:28:52 &  98.1  &  7.9 & (PSPC B)\\
\noalign{\smallskip}
{\it RXTE}     & 14 Dec 1997 & 03:44:11 &   15.0 &  8.4 & (PCA)\\
               &             & 03:44:11 &   15.0 &  2.6 & (HEXTE)\\
\noalign{\smallskip}
{\it BeppoSAX} & 04 May 1998 & 22:05:14 &   92.4 & 20.0 & (LECS)\\
	       &	     & 22:05:14 &   92.4 & 46.4 & (MECS)\\
 	       &	     & 22:05:14 &   92.4 & 20.0 & (HPGSPC)\\
	       &             & 22:05:14 &   92.4 & 21.0 & (PDS)\\
\noalign{\smallskip}
{\it Swift}    & 19 Apr 2006 & 21:02:10 &    7.1 &  2.5 & (XRT)\\
\noalign{\smallskip}
\hline
\noalign{\smallskip}
\multicolumn{6}{l}{$^*$Results from this observation were partially
presented by Cook et al. (1984)}\\
\noalign{\smallskip}
\hline
\hline
\noalign{\smallskip}
\end{tabular}
\end{center}
\end{table*}

X--ray binaries are interacting systems composed of a compact object, 
neutron star (NS) or black hole (BH), which accretes matter from a 
``normal'' star orbiting at close distance. According to the mass of the 
donor star, these systems are classified as High Mass X--Ray Binaries 
(HMXBs) or Low Mass X--ray Binaries (LMXBs). Among LMXBs, the mass transfer 
generally occurs through inner Lagrangian point matter overflow from a Roche 
Lobe filling red dwarf (e.g., White et al. 1995; van Paradjs \& McClintock 
1995).

However, there exists a very small subclass of LMXBs (3 out of more than 
150 objects: Liu et al. 2001; Masetti et al. 2006a) that host an M-type 
giant mass-losing star. This red giant is believed to feed the compact 
companion via its stellar wind. By analogy with the symbiotic stars, which are 
systems composed of an M-type giant and a white dwarf (WD), these LMXBs are 
sometimes referred to as `symbiotic X--ray binaries'. The protoype of this 
small subclass, the system GX 1+4, is a well-studied case (e.g., 
Chakrabarty \& Roche 1997); the second member, object 4U 1700+24, is now
receiving more attention (Masetti et al. 2002, 2006a; Galloway et al. 2002; 
Tiengo et al. 2005).

The third member of this class, 4U 1954+31, despite being relatively 
luminous in X--rays (it is an {\it Uhuru} source), has been neglected over the 
years. The object was first detected in the '70s in this spectral window
by {\it Uhuru} (Forman et al. 1978) and subsequently by {\it Ariel V} 
(Warwick et al. 1981) in the framework of their all-sky surveys. Warwick et 
al. (1981) found the source at an average flux of 
$\approx$2$\times$10$^{-10}$ erg cm$^{-2}$ s$^{-1}$ in the 2--10 
keV band, with long-term variability of a factor of more than 20.

The first pointed observation was performed with {\it EXOSAT} in October 
1983. The data (Cook et al. 1984) showed flaring behavior on 1--10 min 
timescales but no evidence of pulsations or periodic variations; the 
1--20 keV spectrum was modeled with either a powerlaw with photon index 
$\Gamma \sim$ 1.5 or a bremsstrahlung model with $kT \sim$ 36 keV. These 
characteristics led Cook et al. (1984) to conclude that 4U 1954+31 is 
probably an HMXB hosting a compact object accreting from a highly 
inhomogeneous wind from a companion star of early spectral type.

Subsequently, 4U 1954+31 was observed in October 1987 by {\it Ginga} in 
the 2--30 keV band (Tweedy et al. 1989). This observation confirmed the 
timing characteristics found by Cook et al. (1984) and allowed Tweedy et 
al. (1989) to model the spectrum of the source with a high-energy cutoff 
powerlaw plus a soft excess below 4 keV, heavily absorbed by intervening 
matter (possibly local to the source) partially covering the soft X--ray 
emission. These authors also found spectral variability which they 
interpreted as due to changes in both the continuum shape and the absorption 
column. Tweedy et al. (1989) also (unsuccessfully) searched for the optical 
counterpart of this source, concluding that the M-type giant within the 
{\it EXOSAT} error circle is unlikely associated with 4U 1954+31 and that 
the true counterpart should be an H$_\alpha$-emitting, highly reddened 
object (most likely a supergiant secondary hosted in an HMXB).

Notwithstanding all of the above, which makes 4U 1954+31 a potentially 
interesting target for X--ray studies, no refereed publications were 
devoted to this source, and no optical counterpart has been associated with 
it, up to March 2006. After a hiatus of nearly 20 years, 
Masetti et al. (2006a), thanks to a pointed {\it Chandra} observation and a 
spectroscopic optical follow-up, showed that the M-type giant mentioned 
above is indeed the optical counterpart of 4U 1954+31, and that this 
system is a symbiotic X--ray binary located at a distance $d \la$ 1.7 kpc.

The work of Masetti et al. (2006a) shed light on this 
high-energy emitting source, and revived the interest in this object: a 
fast flare (lasting less than 2 hours) from 4U 1954+31 was detected with 
{\it INTEGRAL} in April 2006 and was communicated by Paltani et al. (2006); 
also, Corbet et al. (2006) found a 5-hour periodic modulation in the 
{\it Swift} BAT data acquired between December 2004 and September 2005.
Mattana et al. (2006), using {\it INTEGRAL} observations and a subset of 
the data presented in this work, confirmed the NS spin period and the high 
absorption in the X--ray spectrum.

In the present paper, we collect all of the unpublished archival X--ray 
observations on this system to fully characterize its high-energy behaviour 
and characteristics in both temporal (long- and short-term) and spectral 
scales. Particular attention is paid to the still unexplored parts of 
the X--ray spectrum of 4U 1954+31, i.e. below 2 keV and above 30 keV, 
where the high sensitivity of {\it BeppoSAX} is of paramount 
importance to gauge the source emission.

The aim of this paper is therefore to study the high energy properties 
of 4U 1954+31, and in particular the soft excess seen by Tweedy et al. 
(1989) below 4 keV. With this purpose, we have analyzed the data collected 
on 4U 1954+31 obtained over a 
time span of $\sim$23 years through pointed observations performed with 
various satellites ({\it EXOSAT}, {\it ROSAT}, {\it RXTE}, {\it Swift} and 
{\it BeppoSAX}). We have also retrived the data collected with the {\it 
RXTE} All-Sky Monitor (ASM), in order to have a better view of the source 
long-term X--ray activity.

\section{Observations and data reduction}

Here we describe the X--ray observations of 4U 1954+31 made with five 
different spacecraft. Table 1 reports the log of all the X--ray 
pointed observations presented in this paper.

\subsection{{\it BeppoSAX} data}

This source was observed with the Narrow Field Instruments (NFIs) onboard 
{\it BeppoSAX} (Boella et al. 1997a) on 4 May 1998. The NFIs included the 
Low-Energy Concentrator Spectrometer (LECS, 0.1--10~keV; Parmar et al. 
1997), two Medium-Energy Concentrator Spectrometers (MECS, 1.5--10~keV; 
Boella et al. 1997b), a High Pressure Gas Scintillation Proportional 
Counter (HPGSPC, 4--120~keV; Manzo et al. 1997), and the Phoswich 
Detection System (PDS, 15--300~keV; Frontera et al. 1997). LECS and MECS 
were imaging instruments, whereas HPGSPC and PDS were non-imaging 
collimated detectors with fields of view of about 1$\fdg$0 and 1$\fdg$3, 
respectively. During all pointings the four NFIs worked nominally and the 
source was detected by all of them.

Good NFI data were selected from intervals outside the South Atlantic 
Geomagnetic Anomaly when the elevation angle above the earth limb was 
$>$ $5^{\circ}$, when the instrument functioning was nominal and, for 
LECS events, during spacecraft night time. The SAXDAS 2.0.0 data analysis 
package (Lammers 1997) was used for the extraction and the processing of 
LECS, MECS and HPGSPC data. The PDS data reduction was instead performed 
using XAS version 2.1 (Chiappetti \& Dal Fiume 1997). LECS and MECS data 
were reduced using an extraction radius of 4$'$, centered on the source 
position; before extraction, data from the two MECS units were merged.

Background subtraction for the two imaging instruments was performed using 
standard library files, while the background for the PDS data was evaluated 
from two fields observed during off-source pointing intervals.
Due to the presence of the strong X--ray source Cyg X-1 in one of the
off-source pointings, only the other one was used for the background
subtraction. For the same reason, the background for the HPGSPC data was 
computed through an Earth-occultation technique (Manzo et al. 1997).

Inspection of the MECS data shows the presence of a field X--ray source, 
located at $\sim$21$'$ from 4U 1954+31, which we identify with the dwarf 
nova EY Cyg. By studying its X--ray spectrum (see Appendix) and 
extrapolating it to the ranges covered by the non-imaging instruments 
HPGSPC and PDS, we could determine the degree of contamination of the 
X--ray emission of this source on the hard X--ray spectrum of 4U 1954+31. 
It is found that the X--ray emission of EY Cyg is less than 0.1\% of the 
flux detected from 4U 1954+31 above 7 keV (see Sect. 3.2), thus
negligible. No contamination from EY Cyg is present in the MECS data 
because it appears fully separated from 4U 1954+31 in the observation 
acquired by this imaging instrument. EY Cyg was not observed by the LECS 
as it was outside its field of view.

We also checked for the presence of other contaminating X--ray sources in 
the PDS and HPGSPC fields of view. Only three {\it ROSAT} sources (Voges 
et al. 1999, 2000) were found at the edge of the PDS field of view, at 
distances of more than 35$'$ from 4U 1954+31.
Assuming conservatively a hard spectral photon index $\Gamma$ = 1 for all 
of them, we determined their total flux convolved with the PDS triangular 
response. We found that their contamination in the 15--150 keV band is 
less than 0.9\% of the flux of 4U 1954+31 in this band (see 
Sect. 3.2), and therefore negligible as well.

In the same way, we evaluated the contamination of field X--ray sources
on the PDS background measurement: 5 {\it ROSAT} sources were found in
the off-source field used for background evaluation; it is found that
these contribute less than 0.35\% of the total measured background.
Thus, again in this case, we can neglect this contribution.

\begin{figure*}
\vspace{-4.5cm}
\hspace{-1cm}
\epsfig{figure=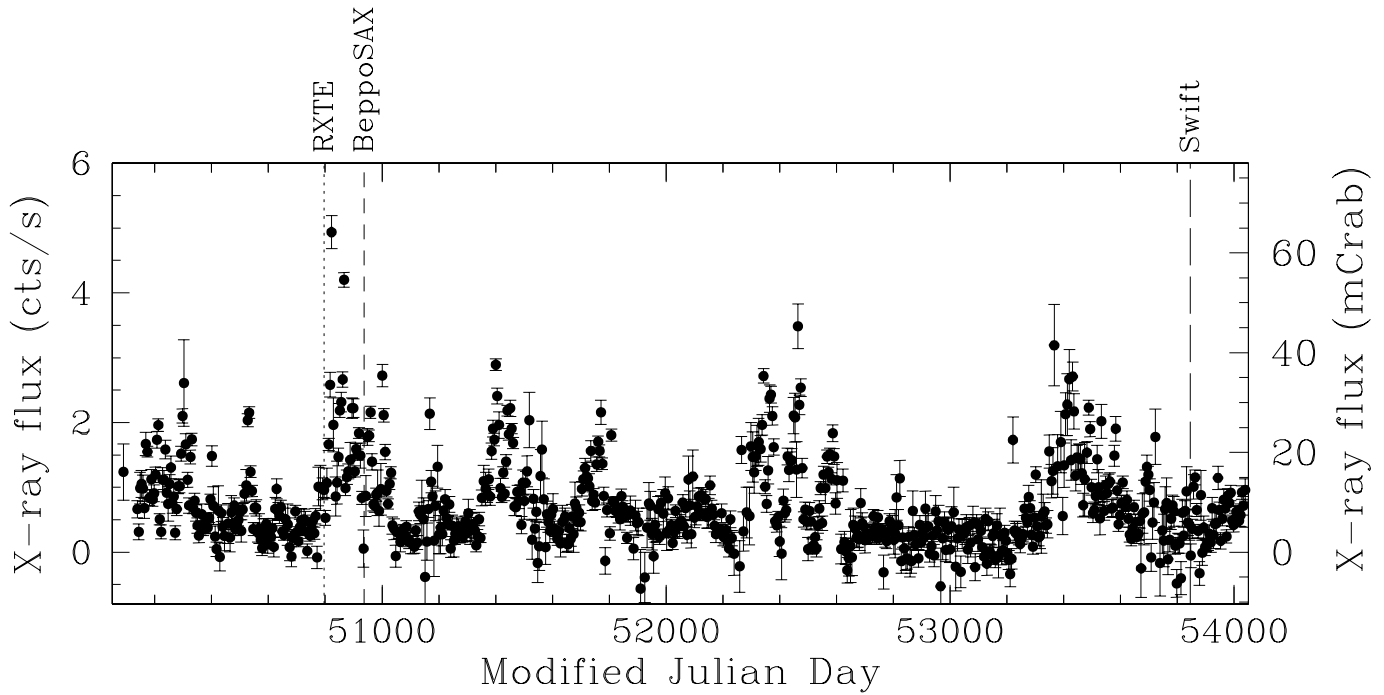,width=14.8cm,angle=270}
\vspace{-1.2cm}
\caption[]{1.5--12 keV 5--day averaged {\it RXTE} ASM light curve of
4U 1954+31. One ASM count s$^{-1}$ roughly corresponds to 13 mCrab assuming
a Crab-like spectrum. In the plot the times of the pointed {\it RXTE},
{\it BeppoSAX} and {\it Swift} are indicated by the vertical dashed lines 
(different dashes correspond to different spacecraft). A series of aperiodic 
increases in the X--ray activity from the source are noticed occurring on 
timescales variable from 200 to 400 days.}
\end{figure*}

\subsection{Other public X--ray data}

In order to present a thorough long-term analysis of the X--ray behaviour 
of 4U 1954+31, we also retrieved from the HEASARC 
archive\footnote{available at:~\texttt{http://heasarc.gsfc.nasa.gov/\\ 
cgi-bin/W3Browse/w3browse.pl}} all the unpublished X--ray observations of 
this source. These include (see Table 1) {\it ROSAT}, {\it RXTE}, {\it 
EXOSAT} and {\it Swift} data; we also re-examine the {\it EXOSAT} pointing 
already partially studied by Cook et al. (1984).

Two pointed observations were performed on 4U 1954+31 with the {\it 
EXOSAT} satellite (White \& Peacock 1988). The first one ({\it EXOSAT} I) 
was acquired on 22 October 1983 while the second one ({\it EXOSAT} II)
on 2 June 1985. Both observations were made using the Medium Energy 
(ME; Turner et al. 1981) proportional counter, in the 1--50 keV band, and 
with the Gas Scintillation Proportional Counter (GSPC; Peacock et al. 
1981), in the range 2--20 keV. However, due to their very low 
signal-to-noise ratio (S/N), GSPC data of {\it EXOSAT} II were not used 
in this analysis.

The {\it ROSAT} satellite (Tr\"umper 1982) was pointed at 4U 1954+31 on 
4 May 1993 with the Position Sensitive Proportional Counter (PSPC; 
Pfeffermann \& Briel 1986) unit B, sensitive in the 0.1--2.4 keV range.

The {\it RXTE} satellite (Bradt et al. 1993) observed the source on 14 
December 1997; this satellite carries a 5-unit Proportional Counter Array 
(PCA; Jahoda et al. 1996), which is sensitive in the 2--60 keV energy 
range and allows a time resolution of 1 $\mu$s, and a High Energy X--ray 
Timing Experiment (HEXTE; Rothschild et al. 1998) composed of two clusters 
of 4 phoswich scintillation detectors working in the 15--250 keV band.

This satellite also carries an ASM\footnote{ASM light curves are available 
at:\\ \texttt{http://xte.mit.edu/ASM\_lc.html}} (Levine et al. 1996) which 
regularly scans the X--ray sky in the 1.5--12 keV range with a daily 
sensitivity of 5 mCrab. Figure 1 reports the complete 1.5--12 keV ASM 
light curve of 4U 1954+31 starting on January 1996 up to October 
2006, along with the times of the {\it BeppoSAX}, {\it RXTE} and {\it 
Swift} pointed observations. In order to clearly display the long-term 
trend in the X--ray emission from 4U 1954+31, each point in Fig. 1 is 
computed as the average of 5 subsequent measurements; given that the 
single original ASM points we retrived were acquired on a daily 
basis (i.e. they are one-day averaged measurements), the plot illustrated 
in Fig. 1 corresponds to a 5-day averaged X--ray light curve.

Further X--ray data of 4U 1954+31 were collected on 19 April 2006 
with the X--Ray Telescope (XRT, 0.2--10 keV; Burrows et al. 2006) 
on board {\it Swift} (Gehrels et al. 2004).
The data reduction was performed using the XRTDAS v2.4 standard data 
pipeline package ({\tt xrtpipeline} v0.10.3), in order to produce the 
final cleaned event files.

As the XRT count rate of the source was high enough to cause pile-up in 
the data, we extracted the source events in an annulus, of 12$''$ inner 
radius and 45$''$ outer radius, centered on the source. The size of the 
inner region was determined following the procedure described in Romano et 
al. (2006). The source background was measured within a circle with radius 
80$''$ located far from the source. The ancillary response file was 
generated with the task {\tt xrtmkarf} (v0.5.2) within 
FTOOLS\footnote{available at:\\
\texttt{http://heasarc.gsfc.nasa.gov/ftools/}} (Blackburn 1995), and 
accounts for both extraction region and PSF pile-up correction. We used 
the latest spectral redistribution matrices in the Calibration 
Database\footnote{available at: 
{\tt http://heasarc.gsfc.nasa.gov/\\docs/heasarc/caldb/caldb\_intro.html}} 
(CALDB 2.3) maintained by HEASARC.

We determined the source position using the {\tt xrtcentroid} (v0.2.7) 
task. The correction for the misalignment between the telescope and the 
satellite optical axis was taken into account (see Moretti et al. 2006 for 
details). The {\it Swift} XRT position we obtained for 4U1954+31 is the 
following (J2000): RA = 19$^{\rm h}$ 55$^{\rm m}$ 42$\fs$28; Dec = 
+32$^\circ$ 05$'$ 45$\farcs$5 (with a 90\% confidence level error of 
3$\farcs$6 on both coordinates). 

These coordinates are consistent with the (more accurate) ones measured in 
X--rays with {\it Chandra} and with those of the optical counterpart 
(Masetti et al. 2006a). We remark that the uncertainty in the source 
coordinates as determined with the {\it Swift} XRT data may be 
underestimated because, as said above, the XRT observation is affected by 
pile-up and the PSF of the source image is thereby distorted.

\section{Results}

\subsection{Light curves and timing analysis}

\begin{figure*}[t!]
\epsfig{figure=6517f2ul.ps,width=6.5cm,angle=270}

\vspace{-7.5cm}
\hspace{8.3cm}
\epsfig{figure=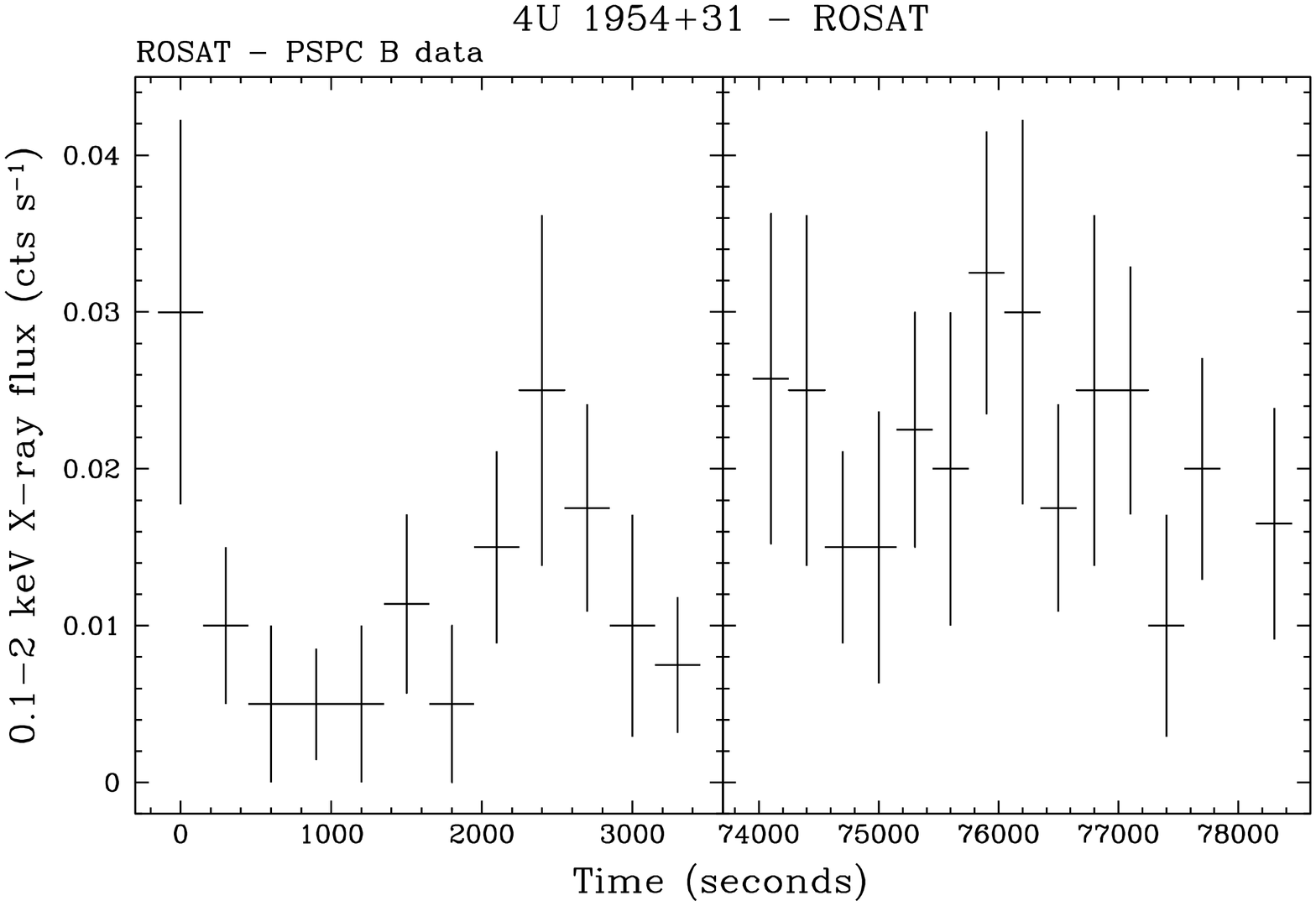,width=8cm,height=10cm,angle=270}

\vspace{-0.2cm}
\epsfig{figure=6517f2cl.ps,width=6.5cm,angle=270}
\hspace{.4cm}
\epsfig{figure=6517f2cr.ps,width=6.5cm,height=8.5cm,angle=270}

\vspace{0.2cm}
\epsfig{figure=6517f2ll.ps,width=6.5cm,height=8.5cm,angle=270}
\hspace{.4cm}
\epsfig{figure=6517f2lr.ps,width=6.5cm,height=8.5cm,angle=270}
\caption{Background-subtracted X--ray light curves of 4U 1954+31 as seen 
during the pointed observations reported in the text. {\it (Upper panels)} 
2--10 keV light curve of {\it BeppoSAX} MECS (left) and 0.1--2 keV light 
curve of {\it ROSAT} PSPC B (right). The former is binned at 200 s, the
latter at 300 s. {\it (Central panels)} 2--9 keV {\it RXTE} PCA (left) and 
1--10 keV {\it Swift} XRT (right) light curves. Both light curves are 
binned at 50 s. {\it (Lower panels)} 0.8--8.9 keV light curves of {\it 
EXOSAT} ME observations I (left) and II (right). The former is binned at 
60 s, the latter at 100 s; for the sake of comparison of the activity of 
the source between the two {\it EXOSAT} observations, the latter panels were
plotted using the same $y$-axis scaling. Given that, during the {\it ROSAT} 
observation, good source events were recorded only at the beginning and at 
the end of the pointing, we plot here these two time intervals. In all 
figures, times are in seconds since the beginning of the observation as 
reported in Table 1.}
\end{figure*}

The {\it RXTE} long-term (January 1996/October 2006) 1.5--12 keV ASM 
light curve (see Fig. 1) of 4U 1954+31 shows an erratic behaviour of the 
source with sudden increases in its activity. The timing analysis of the 5-day 
averaged data has not revealed a coherent periodicity but only a 
variability on timescales between 200 and 400 days. A similar indication 
is obtained using the dwell-by-dwell and the daily-averaged ASM data sets.

We also studied the short-term X--ray light curves obtained with the 
pointed observations on 4U 1954+31 (see Fig. 2). The light curves have 
different bin times (from 50 s to 300 s) depending on their S/N. 

All light curves show an erratic flaring activity on 
timescales variable between 100 s and 1000 s, the longer ones 
corresponding to the more intense flares. In particular, {\it EXOSAT} I 
shows the most intense flaring activity among the observations presented 
here, while during {\it EXOSAT} II the source was more active in the first 
part of the observation.

As Cook et al. (1984) and Tweedy et al. (1989) pointed out some 
variability in the source hardness ratio as a function of the X--ray count 
rate, we investigated the presence of this possible correlation for 4U 
1954+31. However, we did not find any change in the spectral hardness with 
X--ray intensity in any observation. Only {\it RXTE} data suggest a slight 
hardening with increased source intensity.

Timing analysis on the X--ray light curves from the various spacecraft was 
performed with the FTOOLS task XRONOS, version 5.21, after having 
converted the event arrival times to the solar system barycentric frame. 
The data were searched for coherent pulsations or periodicities but no 
evidence for them was found in the range 2 ms to 2000 s, although above 10 
s any coherent behaviour could be significantly masked by the flaring 
activity of 4U 1954+31 (see Fig. 2).
The power spectral densities obtained for each pointing are characterized 
by red noise and show no significant deviation from the 1/$f$-type 
distribution (where $f$ is the time frequency) typical of the `shot-noise' 
behaviour.

Likewise, a Fast Fourier Transform (FFT) analysis was made to search for
fast periodicities (pulsations or QPOs) in the 1--2000 Hz range within the 
{\it RXTE} PCA data.
During the observation, data in ``Good Xenon" mode with a 2$^{-20}$ s
(i.e. 1 $\mu$s) time resolution and 256 energy bands were available.
To search for fast pulsations and/or kiloHertz QPOs we made several 
FFTs of 16 s long data segments in the 2--10 keV and 10--60 keV energy
intervals and with a Nyquist frequency of 4096 Hz.
We then calculated the $Z^2$ statistics (Buccheri et al. 1983) on each 
$\sim$3000 s long subinterval of the {\it RXTE} PCA observation (see Fig. 
2, central left panel) and finally added together the results. No presence 
of QPO peaks or of coherent pulsations was detected.

\subsection{Spectra}

\begin{figure*}[t!]
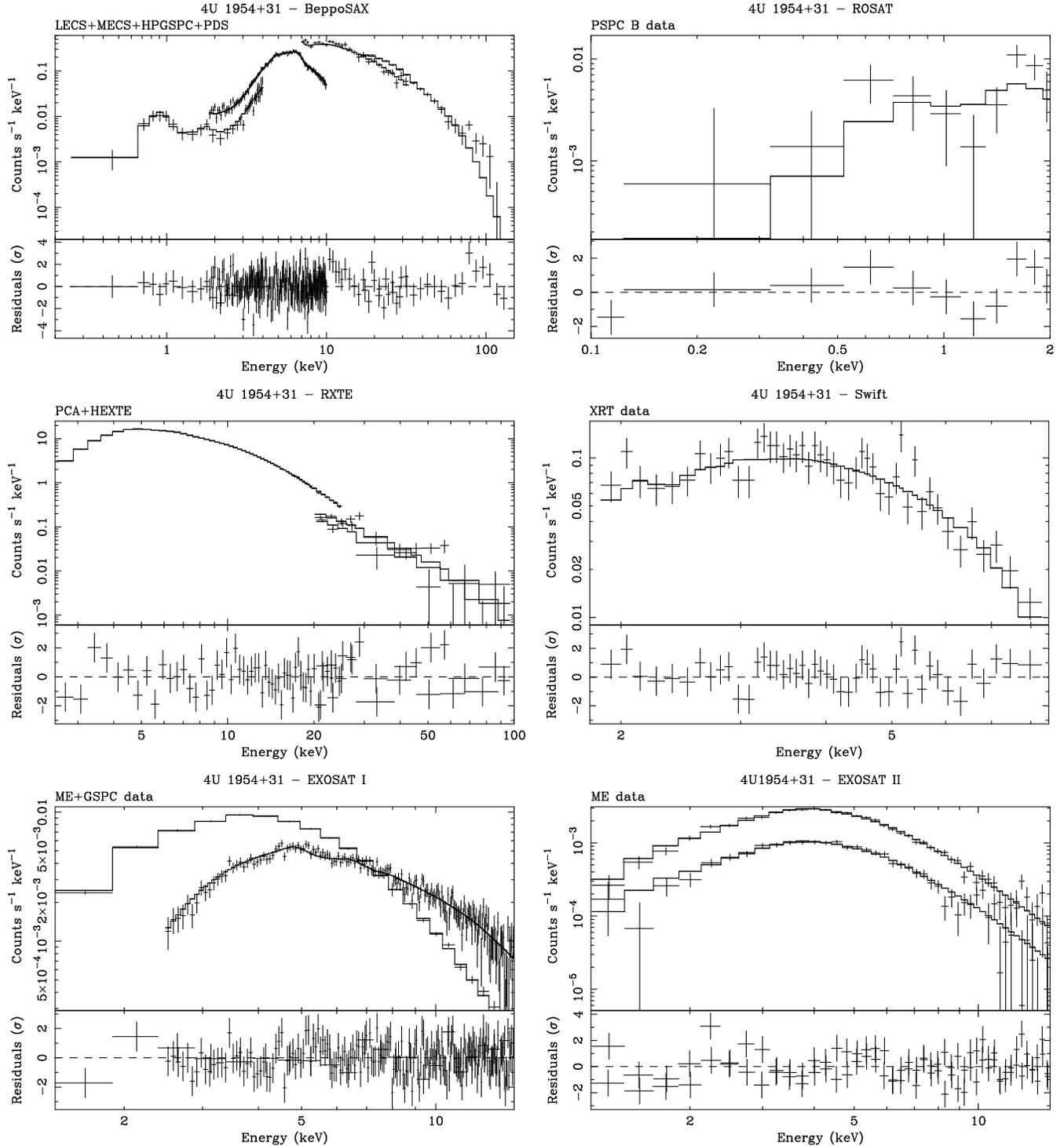

\epsfig{figure=6517f3ul.ps,width=6.3cm,angle=270}
\hspace{.3cm}
\epsfig{figure=6517f3ur.ps,width=6.3cm,angle=270}

\vspace{0.2cm}
\epsfig{figure=6517f3cl.ps,width=6.3cm,angle=270}
\hspace{.1cm}
\epsfig{figure=6517f3cr.ps,width=6.3cm,angle=270}

\vspace{0.2cm}
\epsfig{figure=6517f3ll.ps,width=6.3cm,angle=270}
\hspace{.3cm}
\epsfig{figure=6517f3lr.ps,width=6.3cm,angle=270}
\vspace{-.2cm}
\caption{X--ray spectra of 4U 1954+31 obtained from the pointed 
observations described in the text. The fit residuals using the best-fit 
model reported in Table 2 are also associated with each spectrum. {\it 
(Upper panels)} {\it BeppoSAX} LECS+MECS+HPGSPC+PDS (left) and {\it ROSAT} 
PSPC B (right) spectra. {\it (Central panels)} {\it RXTE} PCA+HEXTE (left) 
and {\it Swift} XRT (right) spectra. {\it (Lower panels)} spectra of {\it 
EXOSAT} observations I (ME+GSPC data; left) and II (ME data; right). For 
observation II, the spectra of the high and low states (first 5 ks and 
remaining part of the observation, respectively; see Fig. 2, upper right 
panel, for reference) are shown.}
\end{figure*}

\begin{table*}
\begin{center}
\caption[]{Best-fit parameters for the X--ray spectra of 4U 1954+31 from
the pointed observations described in this paper. In all models in which a
Fe emission line is needed in the fit, its width is fixed at the value
$\sigma = 0.1$ keV. Frozen parameters are written between square
brackets. Luminosities are not corrected for the intervening absorptions 
and are computed assuming a distance $d$ = 1.7 kpc (Masetti et al. 2006a); 
in the cases in which the X--ray data could not completely cover the 
X--ray interval of interest for the luminosity determination, an 
extrapolation of the best-fit model is applied. Errors are at 90\% 
confidence level for a single parameter of interest. Observations are 
reported in chronological order from left to right.}
\begin{tabular}{c|c|c|c|c|c|c}
\hline
\noalign{\smallskip}
\hline
\noalign{\smallskip}

\multicolumn{7}{c}{model: {\sc wabs*[pcfabs*wndabs(comptt + gauss) + mekal]}} \\

\noalign{\smallskip}
\hline
\noalign{\smallskip}

Parameters$^{\rm a}$ & {\it EXOSAT} I & {\it EXOSAT} II & {\it ROSAT} & {\it RXTE} & {\it BeppoSAX} & 
{\it Swift} \\
 & (1--15 keV) & (1--15 keV) & (0.1--2 keV) & (2.5--100 keV) & (0.2--150 keV) & 
(1.8--8.5 keV) \\
\noalign{\smallskip}
\hline
\noalign{\smallskip}

${\chi^{2}}$/dof         & 217.7/204 & 86.0/85 & 20.0/18 & 112.6/105 & 280.1/247 & 53.4/44 \\


$N_{\rm H}^{\rm WABS}$   & [1.5] & [1.5] & [1.5] & [1.5] & 1.5$^{+0.6}_{-0.4}$ & [1.5] \\

$N_{\rm H}^{\rm PCFABS}$ & --- & --- & [25.4] & --- & 25.4$\pm$1.6 & 1.4$^{+2.6}_{-1.2}$ \\
$C_{\rm PCFABS}$ (\%)    & --- & --- & [95] & --- & 95$\pm$2 & [95] \\

$N_{\rm H}^{\rm WNDABS}$ & --- & --- & [3.0] & 4.1$^{+0.2}_{-0.3}$ & 3.0$^{+1.3}_{-0.7}$ & --- \\
$E_{\rm WNDABS}$ (keV)   & --- & --- & [1.86] & [1.86] & 1.86$^{+0.10}_{-0.13}$ & --- \\

$kT_0$ (keV)             & 0.56$\pm$0.04 & 0.94$^{+0.05}_{-0.06}$ & [1.28] & 
1.156$^{+0.017}_{-0.009}$ & 1.28$\pm$0.06 & 0.6$^{+0.4}_{-0.6}$ \\
$kT_{\rm e^{-}}$ (keV)   & 2.86$^{+0.12}_{-0.10}$ & 13$^{+\infty}_{-9}$ & [9.9] & 13$^{+14}_{-3}$ & 
9.9$\pm$0.8 & 8$^{+\infty}_{-2}$ \\
$\tau$                   & 9.4$^{+0.5}_{-0.4}$ & 3.2$^{+2.5}_{-3.2}$ & [2.8] & 2.0$^{+0.5}_{-0.4}$ & 
2.8$^{+0.3}_{-0.2}$ & 4$^{+5}_{-4}$ \\
$K_{\rm COMPTT}$ ($\times$10$^{-3}$) & 50$^{+2}_{-3}$ & 2.1$^{+3.6}_{-2.1}$ & [8.2] & 
5.0$^{+0.4}_{-2.4}$ & 8.2$\pm$0.8 & 6$^{+120}_{-6}$ \\

$kT_{\rm MEKAL}$ ($\times$10$^{-2}$ keV) & --- & --- & 5.3$^{+1.5}_{-1.2}$ & --- & 6.8$^{+2.2}_{-1.5}$ & --- \\
$K_{\rm MEKAL}$          & --- & --- & 300$^{+5600}_{-290}$ & --- & 200$^{+5400}_{-190}$ & --- \\

$E_{\rm Fe}$ (keV)       & --- & --- & N/A & 6.46$^{+0.12}_{-0.11}$ & 6.46$^{+0.10}_{-0.11}$ & --- \\
$I_{\rm Fe}$ ($\times$10$^{-4}$ ph cm$^{-2}$ s$^{-1}$) & --- & --- & N/A & 1.9$^{+0.5}_{-0.4}$ & 
2.5$\pm$1.0 & --- \\

\noalign{\smallskip}
\hline
\noalign{\smallskip}
\multicolumn{1}{l|}{X--ray luminosities$^{\rm b}$:} & & & & & & \\

0.1--2 keV  & --- & --- & 4.1$\times$10$^{31}$ & --- & 2.0$\times$10$^{32}$ & --- \\
2--10 keV   & 1.8$\times$10$^{35}$ & 3.4$\times$10$^{34}$ & --- & 7.9$\times$10$^{34}$ & 
5.2$\times$10$^{34}$ & 5.2$\times$10$^{34}$ \\
10--100 keV  & --- & --- & --- & 1.2$\times$10$^{35}$ & 1.8$\times$10$^{35}$ & --- \\
\noalign{\smallskip}
\hline
\noalign{\smallskip}
\multicolumn{7}{l}{$^{\rm a}$Hydrogen column densities are in units of 10$^{22}$ cm$^{-2}$.} \\
\multicolumn{7}{l}{$^{\rm b}$In units of erg s$^{-1}$. Values are not corrected for absorption.} \\
\multicolumn{7}{l}{N/A: not applicable because of lack of spectral coverage.} \\
\noalign{\smallskip}
\hline
\hline
\end{tabular}
\end{center}
\end{table*}

In order to perform spectral analysis, the pulse-height spectra from the
detectors of all spacecraft were rebinned to oversample by a factor 3
the full width at half maximum of the energy resolution and to have a 
minimum of 20 counts per bin, such that the $\chi^2$ statistics could 
reliably be used.
For all detectors, data were then selected, when a sufficient number of
counts were obtained, in the energy ranges where the instrument responses
are well determined.
In all cases we considered the average spectra of 4U 1954+31 given that 
no substantial variations in the spectral shape during each pointed 
observation were suggested by the inspection of the color-intensity 
diagram of the source (See Sect. 3.1).

Spectral analysis was performed with the package {\sc xspec} (Arnaud 1996) 
v11.3.1. In the broad band {\it BeppoSAX} fits, 
normalization factors were applied to LECS, HPGSPC and PDS spectra following 
the cross-calibration tests between these instruments and the MECS: these 
factors were fixed to the best estimated values (Fiore et al. 1999). 
Similarly, a constant was introduced between HEXTE and PCA in the {\it RXTE} 
spectra as well as between ME and GSPC in the {\it EXOSAT} spectra to take 
the different instrumental sensitivities into account. 
The spectra from the two HEXTE clusters were fitted independently and the 
normalization between them was left free, because of the presence of a 
small systematic difference between their responses. 

To all spectral models tested in this paper (the best-fit models for each 
satellite pointing are reported in Table 2), we applied a photoelectric 
absorption column, modeled using the cross sections of Morrison \& 
McCammon (1983; {\sc wabs} in {\sc xspec} notation), to describe the 
line-of-sight Galactic neutral hydrogen absorption towards 4U 1954+31.
In the following, the acronym `dof' means `degrees of freedom'.
Table 3 reports values or 90\% confidence level upper limits of the 
Equivalent Width (EW) of any Fe emission line at $\sim$6.5 keV
in the spectrum of each pointed observation.

We first considered the 0.2--150 keV {\it BeppoSAX} spectrum as it 
is the one with the widest spectral coverage. The initial attempt to fit 
this spectrum with a simple model such as a power law was unsatisfactory 
and indicated the presence of a soft excess below $\sim$2 keV and of a 
spectral turnover above 20 keV. The use of single-component models 
therefore proved to be inadequate, giving unacceptable reduced $\chi^2$ 
values ($>$3). 

Mattana et al. (2006) described the X--ray spectrum of this source 
above 18 keV with a phenomenological model made of a cutoff power law.
Our 15--150 keV PDS spectrum can be well fit with the same model
and our results are compatible with theirs.
When including the lower part of their broadband X--ray data, Mattana 
et al. (2006) used an absorbed cutoff power law plus blackbody model; 
however, they note that their spectral description is poor at low 
energies. We reached the same conclusion when applying this 
same model to our {\it BeppoSAX} spectrum, which means that a 
different and more physical modeling should be used. 

We therefore assumed, as already applied to the X--ray spectra of this 
type of source (Masetti et al. 2002), a composite model made of a 
Comptonization component ({\sc comptt} in {\sc xspec}; Titarchuk 1994) and 
a thermal component to describe the emissions above and below 2 keV, 
respectively. We however found that, in order to describe the {\it 
BeppoSAX} spectrum at low energy, a complex absorption pattern was needed, 
with different absorption columns acting on the two components. In 
particular we found that, while on the thermal component a simple neutral 
hydrogen absorbing column was needed, a mixture of a partial covering 
absorber and of an ionized absorber (as modeled by Baluci\'{n}ska-Church 
\& McCammon 1992) was found acting on the Comptonization component.
We also noted that the thermal component could be described with an
optically thin hot diffuse gas ({\sc mekal}; Mewe et al. 1985)
at relatively low temperature ($kT_{\rm MEKAL} \sim$ 0.07 keV); the 
thermal component could not be fitted with a blackbody or a disk-blackbody
emission.

The spectral fit is slightly improved ($\Delta \chi^2$ = 4 for 2 dof less) 
by the addition of a narrow (with width $\sigma$ fixed at 0.1 keV) Fe 
emission line at $\sim$6.5 keV.
In order to estimate the significance of this line, we used the F-test
statistics for the case of an additional component (e.g., Bevington 1969): 
we find that the line has a chance improvement probability of
1$\times$10$^{-3}$. The {\it BeppoSAX} spectrum and the corresponding 
best fit model are reported in Fig. 3, upper left panel.

To further study the presence of the soft excess found in the {\it 
BeppoSAX} spectrum, we analyzed the 0.1--2 keV spectral data of the {\it 
ROSAT} observation. Due to the fact that the {\it ROSAT} spectrum has a 
smaller energy coverage and suffers from a lower S/N than the 
{\it BeppoSAX} one, we had to constrain the absorption and Comptonization 
component parameters to the {\it BeppoSAX} best-fit values in order to 
study the optically thin plasma emission with a reasonable degree of 
precision. We find that the results (Fig. 3, upper right panel) are 
consistent with those of {\it BeppoSAX}. It should be noted that an 
acceptable description of the spectrum (with $\chi^{2}$/dof = 17.7/18) 
could also be achieved without adding the partially covering absorption 
column; the parameters describing the low-energy thermal emission are 
in this latter case practically the same as those of the model with the 
partial absorption and reported in Table 2.

The analysis of {\it RXTE}, {\it Swift} and {\it EXOSAT} spectra
can instead allow one to monitor the behaviour of the Comptonization
component and of the absorption column associated with it as a function
of time and of X--ray luminosity of the source.
Using the model that best fits the {\it BeppoSAX} data we thus tried
to describe the spectra of 4U 1954+31 secured with these X--ray 
satellites. However, as they do not cover the low-energy extreme
of the X--ray band covered with {\it BeppoSAX} (i.e., below $\sim$1--2 
keV), they are insensitive to the presence of the {\sc mekal} and to the
Galactic line-of-sight hydrogen column. Thus, we modeled these spectra 
with an absorbed Comptonization model alone, and we fixed the Galactic 
$N_{\rm H}$ associated with the {\sc wabs} component to the best-fit 
{\it BeppoSAX} value. 

The {\it RXTE} PCA+HEXTE spectrum, reported in Fig. 3, central left panel, 
apparently does not need the presence of a partial convering absorption, 
and only a warm absorber is required. However, the fit is not formally 
acceptable (with $\chi^2$/dof = 170.0/107) and shows large residuals 
around 6.5 keV. We therefore added a narrow emission line similar to that 
of the {\it BeppoSAX} best-fit model. This new description gives a 
$\Delta\chi^2$ = 57.4 for 2 dof less, meaning a chance improvement 
probability of 4$\times$10$^{-10}$ according to the F-test statistics. 
The best-fit parameter values of the Comptonization component as 
determined from the {\it RXTE} data are comparable to those measured from 
the {\it BeppoSAX} spectrum (see Table 2).

{\it EXOSAT} pointings caught 4U 1954+31 at different emission levels, 
with the source being on average six times brighter during observation I 
than in observation II. During both pointings, the spectral description 
does not need any absorption besides the Galactic one. It is found that, 
while observation II data give Comptonization values broadly similar to 
those of the {\it BeppoSAX} observation (which was performed at a similar 
2--10 keV source flux), observation I shows that 4U 1954+31 had a larger 
Compton cloud optical depth $\tau$ and smaller electron cloud and seed 
photon temperatures $kT_{e^-}$ and $kT_0$. We also find that the 
addition of an emission line at 6.5 keV is not needed in the best-fit 
model for both observations, with EW upper limits slightly tighter than, 
but broadly consistent with, those obtained by Gottwald et al. (1995) from 
the same data. The spectra of the {\it EXOSAT} pointings are reported in 
the lower panels of Fig. 3.

{\it Swift} XRT data (Fig. 3, central right panel) span a smaller spectral 
range and have a lower S/N with respect to the {\it BeppoSAX}, {\it RXTE} 
and {\it EXOSAT} observations; nevertheless, the Comptonization model fits 
the {\it Swift} spectrum reasonably well, with parameter values (although 
affected by large errors) midway between those of the {\it BeppoSAX} and 
{\it EXOSAT} I best fits. The best-fit {\it Swift} spectrum does not need 
the presence of either an ionized absorber, or an iron emission line at 
6.5 keV, and has a lower partially covering absorption column density with 
respect to the {\it BeppoSAX} observation.

We again explored the high-energy part of the {\it BeppoSAX} spectrum to 
look for the presence of a Cyclotron Resonant Feature (CRF) in absorption, 
which is considered one of the signatures of a highly magnetic NS 
hosted in an X--ray binary system (e.g., Orlandini \& Dal Fiume 2001 and 
references therein). We used a {\sc cyclabs} multiplicative model (Mihara 
et al. 1990; Makishima et al. 1990), but it did not statistically improve 
the best-fit model of Table 2. 
Among the other pointings, thanks to its spectral coverage, only {\it RXTE} 
potentially allows one to perform a similar study. However, the lower S/N 
of HEXTE data did not allow us to explore the {\it RXTE} spectra.

As a further test of the presence of a CRF, we also performed a Crab 
ratio of the {\it BeppoSAX} 
PDS spectrum. This technique was used to successfully pinpoint CRFs in 
numerous X--ray binary pulsars (see, e.g., Orlandini et al. 1998), and 
consists of obtaining the ratio between the count rate spectrum of the 
source and that of the featureless power law spectrum of the Crab. This 
ratio has the advantage of minimizing the effects due to detector 
response, non-perfect modeling of the spectral continuum and calibration 
uncertainties. Any real feature present in the source spectrum can then be 
enhanced by multiplying the ratio by the functional form of the Crab 
spectrum (a power law with photon index $\Gamma$ = 2.1) and by dividing it 
by the functional form of the continuum adopted to fit the broad-band 
source spectrum, obtaining the so-called Normalized Crab Ratio (NCR; see 
Orlandini et al. 1998 for details).

As a result, the NCR performed on the PDS spectrum of 4U 1954+31 further 
excludes the presence of any statistically significant absorption feature 
in the 15--150 keV range. This is at odds with the estimate of the magnetic 
field ($B \sim$ 10$^{12}$ Gauss) of the NS hosted in this system made by 
Mattana et al. (2006), as it would produce a clear CRF around 25 keV.
The 90\% confidence level upper limit of the depth of any CRF
at this energy is $<$0.25 assuming a feature width of 10 keV.

\begin{table}
\caption[]{List of the 6.5 keV Fe emission line EWs, assuming a fixed line 
width of 0.1 keV, for the various X--ray observations covering that 
spectral energy. Errors and upper limits are at a 90\% confidence level. 
Pointings are listed in chronological order from top to bottom.}
\begin{center}
\begin{tabular}{c|c}
\noalign{\smallskip}
\hline
\hline
\noalign{\smallskip}
Observation & \textit{EW} (eV) \\
\noalign{\smallskip}
\hline
\noalign{\smallskip}
{\it EXOSAT} I         & $<$56 \\
{\it EXOSAT} II        & $<$73  \\
{\it RXTE}             & 48$^{+13}_{-10}$ \\
{\it BeppoSAX}         & 51$\pm$20 \\
{\it Swift}            & $<$1100 \\
\noalign{\smallskip}
\hline
\hline
\end{tabular}
\end{center}
\end{table}

\section{Discussion}

The broadband, long-term X--ray study of 4U 1954+31 presented here, and 
performed through the analysis of data collected by five satellites, gave 
us the possibility to thoroughly study the long- and short-term  
high-energy behaviours of this bright but neglected source. 
In particular, with the widest ever broadband data coverage (0.2--150 keV) 
of this source, secured with {\it BeppoSAX} and shown here for the 
first time, we could characterize the X--ray spectrum of this object
and check the best-fit model thus obtained against the observations 
acquired with {\it RXTE}, {\it ROSAT}, {\it EXOSAT} and {\it Swift}.
Here we will discuss the results of the previous section and will compare 
them with the high energy behaviour of its twin, the symbiotic X--ray 
binary 4U 1700+24.

\subsection{Broadband X--ray properties of 4U 1954+31}

The temporal analysis of the light curves of the {\it RXTE} ASM and of 
the pointed observations confirms that the source has pronounced long-
and short-term variability, but also that it does not show any periodicity
on either small (seconds to tens of minutes) or large (days to years)
timescales. In particular, we did not find evidence for the 5-hour
periodic signal detected by Corbet et al. (2006) and Mattana et al. 
(2006) in any of the satellite pointings. However, as can be seen from 
Fig. 2, this can be due to several factors: (a) the relatively short 
duration of the observations (in general lasting less than one single 
periodicity cycle); (b) the sparse light curve sampling; and (c) the 
presence of large flares throughout each pointing.

The presence of shot-noise stochastic variability (first found by Cook et 
al. 1984) is instead confirmed for the X--ray emission of 4U 1954+31, 
while no short-period hardness ratio variations are detected during each 
pointed observation. These ``colorless" short-term fluctuations in the 
X--ray flux may be produced by instabilities in the accretion 
process, or by inhomogeneities in the velocity and/or density of an 
accreted stellar wind (or by both; e.g., Kaper et al. 1993). Thus, the 
observed random variability points to an explanation for this X--ray 
activity as due to an inhomogeneous accretion flow onto a compact object 
(e.g. van der Klis 1995).

The lack of evidence for X--ray eclipses in 4U 1954+31 suggests that 
either this system is viewed at a low (or intermediate) inclination 
angle, or that the orbital period is quite long (hundreds of days). In 
fact, a hint of variations of the order of 200--400 days is found in the 
secular ASM X--ray light curve. 

Concerning the spectral properties of 4U 1954+31, one can see that the 
best-fit spectral model is typical of accreting systems hosting a NS 
(e.g., Paizis et al. 2006 and references therein). We cannot 
exclude the presence of a BH in this system; however, the temperatures 
associated with the Comptonization component are those generally seen in 
LMXBs with an accreting NS. The detection of a 5-hour modulation 
(Corbet et al. 2006), interpreted as the spin of the accreting object, 
supports the accreting NS scenario.

The presence of a WD in accretion is basically ruled out in terms of both 
X--ray luminosity and spectral shape (de Martino et al. 2004; Suleimanov 
et al. 2005; Barlow et al. 2006). The X--ray emission from 4U 1954+31 is 
also much harder and stronger than that expected from the corona of a 
late-type giant star, this also being generally 4--5 orders of magnitude 
lower (H\"unsch et al. 1998) than the one detected in the pointed 
observations presented here.

As the source X--ray luminosity increases, the 
emission from the object becomes harder. This is consistent with a model 
in which enhancement of accreting matter onto a compact object is 
responsible for the observed spectral shape and variability. In addition, 
the hardening of the spectrum with increasing luminosity seems to indicate 
accretion of matter with low specific angular momentum (i.e., from a wind), 
as suggested by e.g. Smith et al. (2002) and Wu et al. (2002) to explain a 
similar spectral behaviour observed in other X--ray binaries.

Comparison of X--ray spectral properties of the source among the different 
satellite pointings shows a trend in the Comptonization parameters. In 
particular, an increase of $\tau$ and a decrease of $kT_0$ and $kT_{\rm 
e^-}$ is found when the 2--10 keV X--ray luminosity increases above 
10$^{35}$ erg s$^{-1}$. At this luminosity, the source possibly undergoes 
a spectral state change.

We found indications of complex and variable absorption around this X--ray 
source. A hint of this was already qualitatively suggested by Tweedy et 
al. (1989), although their narrower spectral coverage did not allow them 
to explore this behaviour in detail.
Indeed, the upper limit to the interstellar Galactic absorption along the 
source line of sight is estimated as $N_{\rm H}^{\rm Gal}$ = 
0.84$\times$10$^{22}$ cm$^{-2}$ (Dickey \& Lockman 1990); from the {\it 
BeppoSAX} data we find that the line-of-sight hydrogen column is at least 
about a factor of two higher. The presence of a complex, variable and 
stratified neutral plus ionized absorbing medium may be explained by a 
NS accreting from a wind.

Complex absorbers are sometimes seen at energies less than 3 keV and can 
be described with a ``two-zone'' model (Haberl et al. 1989), where 
X--rays pass in series through a highly ionized absorber closer to the 
accretor and then a cold absorber farther away; 4U 1954+31 thus seems 
to belong to this source typology. This spectral behaviour is 
not unexpected from a source embedded in a stellar wind.
In this sense, the detection of a relatively faint and narrow 
iron emission line suggests the presence of dense, ionized and radially 
infalling material in the close vicinity of the X--ray source. 

Likewise, the nondetection of an accretion disk is explained by the low 
angular momentum of the accreted stellar wind, which then falls nearly 
directly onto the NS surface without forming a large accretion disk 
structure. If the NS is magnetized as suggested by the 
periodicity detected by Corbet et al. (2006), the accretion flow in the 
vicinity of the compact object is likely to be funneled onto the 
NS magnetic polar caps, thus leaving little room for the formation of a
disk structure, albeit small. 

Concerning the magnetic field strength of 
the NS hosted in 4U 1954+31 we note that the nondetection of a CRF from
this source is not unexpected, as other long spin period NSs do not show 
unambiguous CRF signatures (e.g., 4U 0114+65; Masetti et al. 2006b).
The absence of a detectable CRF also casts some doubt on the possibility
that this system hosts a highly-magnetized ($B \sim$ 10$^{12}$ Gauss) NS 
as suggested by Mattana et al. (2006). 

A further indication that the accreting matter is flowing onto the NS 
polar caps via magnetic field confinement comes from the estimate of the
size $r_0$ of the region emitting the Comptonization soft X--ray seed 
photons. Following the prescription by in 't Zand et al. (1999) for the
computation of $r_0$, and using the best-fit spectral parameters reported 
in Table 2, we obtain that $r_0 \sim$ 0.6 -- 1.6 km, depending on 
the Comptonization state of the source. This estimate suggests that the 
area emitting soft seed photons on the NS covers only a fraction of its 
surface and it is comparable with the size of the base of an accretion 
column, which is $\approx$0.1 times the NS radius (e.g., Hickox et al. 
2004).

From the above spectral information, we propose a model for this system
similar to that developed by Hickox et al. (2004) for HMXBs hosting a NS 
accreting from the stellar wind of the companion. These authors indeed 
found that a soft excess around 0.1 keV is a ubiquitous feature in the 
X--ray spectrum of HMXBs hosting a pulsating NS. Although the nature of 
the secondary is different in the case of 4U 1954+31, the proposed 
accretion mode (via dense stellar wind) is the same, therefore producing 
a similar behavior in the X--ray emission properties.

Hickox et al. (2004) suggest that, for sources with luminosity $L_{\rm X} 
\la$ 10$^{36}$ erg s$^{-1}$ (see right panel of their Fig. 3), the NS is 
embedded in a diffuse cloud. The inner parts of this cloud may be ionized, 
absorb the blackbody radiation photons from the NS surface and upscatter 
part of them via Comptonization mechanism. The (much cooler) external 
parts of this cloud also emit thin thermal plasma radiation, by 
collisional energization or reprocessing of a fraction of the hard X--ray 
emission coming from the Comptonization region inside the cloud by the 
optically thin external layers. The cloud is also responsible for the 
overall two-zone variable absorption detected in X--rays.

In the light of all of these findings from our X--ray monitoring campaign, 
we conclude that 4U 1954+31 is a symbiotic X--ray binary in which 
a mildly magnetized NS is embedded in, and accreting from, a complex and 
variable stellar wind coming from the secondary star, an M-type giant.

\subsection{Comparison with the system 4U 1700+24}

We now draw a comparison between 4U 1954+31 and the other symbiotic 
X--ray binary 4U 1700+24.

In terms of time scale variations of the X--ray emission, both systems
show fast shot-noise variations, indicating that they are both
powered by accretion onto a compact object, most likely a NS although 
up to now no evidence of a variability associated with a spin period
was found in 4U 1700+24 (see Masetti et al. 2002 for details and an 
interpretation thereof in terms of system geometry). Galloway et al. 
(2002) suggested that the spin nondetection may be due to the fact
that the NS could be slowly rotating. If true, this would be a
further similarity between the two systems.

Regarding the long-term variations, the behaviour seen in the ASM light 
curve of 4U 1954+31 is reminiscent of that shown by 4U 1700+24 and is
interpreted as a modulation linked to the orbital motion of the system 
components (Masetti et al. 2002; Galloway et al. 2002). For 4U 1954+31, 
however, we did not find variations which are as periodic and stable as 
those found in 4U 1700+24 (although see Mattana et al. 2006 for the
possible presence of a 385-day period in 4U 1954+31).

Spectral analysis of these two sources caught at different emission states 
shows that the Comptonization emission parameters of 4U 1954+31, when 
observed at low luminosity (few 10$^{34}$ erg s$^{-1}$ in the 2--10 keV 
band), are comparable to those shown by 4U 1700+24 during the phases of 
enhanced X--ray emission (when it reaches a similar X--ray luminosity, 
i.e., $\sim$10$^{34}$ erg s$^{-1}$; Masetti et al. 2002). So it seems that 
the Comptonization emission mechanism in these two sources forms a 
continuum depending on the luminosity state, thus suggesting a 
comparable overall accretion mechanism for these two X--ray binaries.

The relatively high luminosity of 4U 1954+31, when compared to 4U 1700+24 
(which is on average at luminosities which are 1--2 orders of magnitude 
fainter; Masetti et al. 2002), may indicate an intense stellar wind 
activity or, more likely, that the NS orbits much closer to the M-type 
giant than in the 4U 1700+24 case. This would also explain the
greater absorption in the accreting flow, as in this way the NS hosted 
in 4U 1954+31 would move into a denser medium.

4U 1954+31, differently from 4U 1700+24, does not show the presence of 
blackbody emission from the accreting NS in its X--ray spectrum. This is 
reasonably explained by the fact that the former source shows the presence 
of a complex (and possibly variable) absorption around the X--ray emitter,
which either absorbs the direct NS surface emission (at lower X--ray 
luminosities, when the accretion flow appears to be denser) or upscatters 
it via Comptonization processes (during the higher X--ray luminosity 
states, when the Compton cloud has a greater optical depth $\tau$).

The two systems may however show a similar behaviour concerning the soft 
X--ray excess detected in both of them. Indeed, Tiengo et al. (2005) found 
in the {\it XMM-Newton} spectrum of 4U 1700+24 an excess which they modeled 
with a wide Gaussian line centered on 0.5 keV. A similar feature is present 
in the {\it BeppoSAX} spectrum of 4U 1954+31 presented here, and which we 
modeled as an optically thin plasma emission. It may thus be possible
that the soft excess seen in 4U 1700+24 can be modeled in the same way.

The absence of emission lines (or more generally of features connected with 
the accretion process) in the spectrum of the optical counterpart of 4U 
1954+31 (Masetti et al. 2006a), as in the case of 4U 1700+24, is naturally 
explained by the comparison between the X--ray luminosity of 4U 1954+31 
($\approx$10$^{35}$ erg s$^{-1}$; see Table 2) and the total luminosity of 
its M-type giant companion, $\sim$900 $L_\odot$ (Lang 1992), i.e. 
3.4$\times$10$^{36}$ erg s$^{-1}$, for a M4 III star (Masetti et al. 2006a). 
As most of the luminosity of the M-type giant is emitted in the optical and 
near-infrared bands, it can be understood that X--ray irradiation has little 
effect on the secondary star.

We speculate that the 4U 1954+31 system evolution is halfway 
between that of 4U 1700+24, where the compact object is accreting from a 
wind with lower density (probably because it is orbiting farther away from 
the secondary), and GX 1+4, where a high accretion rate onto the NS is 
apparent from X--ray and optical data (Chakrabarty \& Roche 1997) implying 
that the latter system is tighter (with the secondary filling its Roche 
lobe, thus producing an accretion disk around the compact object) than 4U 
1700+24 and 4U 1954+31.

\section{Conclusions}

We have presented a broadband multi-spacecraft analysis of the 
symbiotic X--ray binary 4U 1954+31. This allowed us, for the first 
time, to explore in detail the X--ray characteristics of this `forgotten' 
source. From the timing analysis we found that, while random X--ray 
variability is present on long and short time scales, no periodic 
modulation is found. X--ray spectroscopy shows that the high-energy 
photons radiated from this source are produced by Comptonization plus thin 
thermal plasma emission. These two components, in particular the 
Comptonization, suffer from complex variable absorption, likely 
originating in the accretion flow itself. An iron emission line is also 
possibly detected.

From the interpretation of the above data and from the comparison with 
similar sources, the scenario that we envisage is that of a mildly 
magnetized NS orbiting around an M-type giant and accreting matter from 
the companion star stellar wind, which is inhomogeneous in terms of 
density and ionization degree, with the inner parts of the radial 
accretion flow possibly ionized by the NS high-energy radiation. We thus 
conclude that 4U 1954+31 is a symbiotic LMXB which has several of the 
characteristics, in terms of accretion structure, usually found in HMXBs. 
This likely led earlier studies to the conclusion (from the X--ray data 
analysis alone) that this source is an HMXB; only recently, thanks to 
combined optical and X--ray observations, was it found that 4U 1954+31 is 
actually an LMXB (Masetti et al. 2006a).

The soft excess observed in the X--ray spectrum of 4U 1954+31 may be in 
the form of a group of discrete X--ray 
emission lines produced by light metals, rather than a continuum emission. 
This is suggested by the fact that this excess is modeled with an optically 
thin thermal plasma model. Unfortunately, the relatively low spectral 
resolution of the data considered in this paper does not allow us to 
better explore this issue. Detailed modeling is needed to 
understand the nature of the observed complex spectral shape below 2 keV. 
To this aim, high-resolution spectroscopy with {\it XMM-Newton} 
or {\it Chandra} is, at present, the optimal choice.

\begin{acknowledgements}
{\it BeppoSAX} was a joint program of ASI and of the Netherlands Agency 
for Aerospace Programs (NIVR).
This research has made wide use of data obtained through the High Energy
Astrophysics Science Archive Research Center Online Service, provided by
the NASA/Goddard Space Flight Center.
This research has also made use of the SIMBAD database, operated at CDS,
Strasbourg, France and of the ASI Space Data Center {\it Swift} archive.
ASM data were provided by the {\it RXTE} ASM teams at MIT and at the 
{\it RXTE} SOF and GOF at NASA's GSFC. 
This work was partially supported through ASI/INAF grant No. I/023/05/0.
We thank the anonymous referee for useful comments. 
\end{acknowledgements}

\appendix

\section{The X--ray emission from the field source EY Cyg}

As explained in Sect. 2.1, the Cataclysmic Variable EY Cyg
is present in the MECS field as a faint X--ray source. In order to 
check for any possible contamination from this object in the HPGSPC 
and PDS data, we extracted and analyzed the available information on EY 
Cyg as observed by the {\it BeppoSAX} MECS. 
Data were extracted from a region of 3$'$ centered on the source; their 
reduction was similar to that applied in Sect. 2.1 to the 4U 1954+31
MECS data, with the difference that, for the spectral analysis, we 
used the appropriate ancillary response function and background data 
for off-axis observations.

The 2--10 keV light curve during the entire {\it BeppoSAX} observation 
showed that EY Cyg had a steady behaviour, with a possible flare at
half of the observation and lasting less than one hour (see Fig. A.1).
The 5--10 keV / 2--5 keV hardness ratio of the source does not show
significant variations during the {\it BeppoSAX} pointing: we therefore 
considered the 2--10 keV spectrum averaged over the entire observation.

\begin{figure}[h!]
\hspace{-.3cm}
\psfig{file=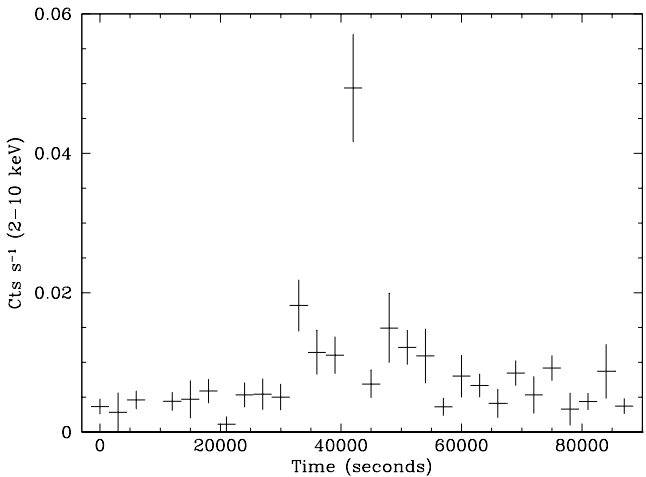,width=7.3cm,angle=-90}
\vspace{-.5cm}
\caption{MECS 2--10 keV countrate light curve of the source EY Cyg.}
\end{figure}

\begin{figure}[h!]
\psfig{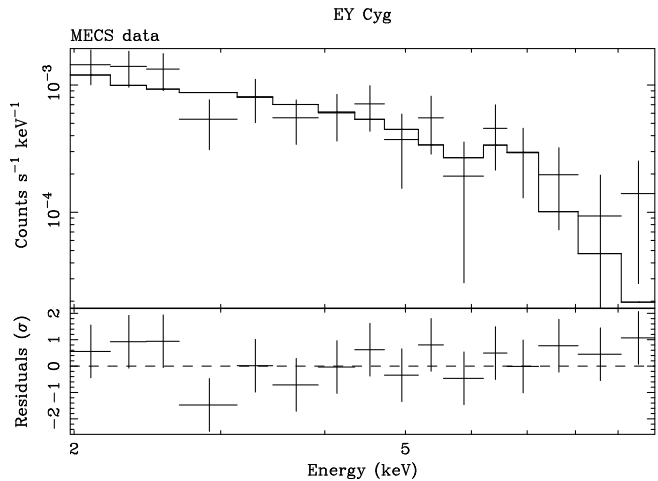}
\caption{MECS 2--10 keV spectrum of the source EY Cyg, averaged over the 
{\it BeppoSAX} observation. Data (in the upper panel) are well fitted 
using a single-temperature optically thin thermal plasma model with 
$kT \sim$ 8 keV. In the lower panel, the best-fit residuals are shown.}
\end{figure}

\begin{table}[h!]
\caption[]{Comparison of the X--ray fluxes (expressed in erg cm$^{-2}$ 
s$^{-1}$) of EY Cyg and 4U 1954+31 in the 7--30 keV and in the 15--150 
keV bands, together with their percentage ratio, as measured during 
the {\it BeppoSAX} pointing described in Sect. 2.1.}
\begin{center}
\begin{tabular}{lccl}
\noalign{\smallskip}
\hline
\hline
\noalign{\smallskip}
\multicolumn{1}{c}{Spectral} & EY Cyg & 4U 1954+31 & \multicolumn{1}{c}{Ratio} \\
\multicolumn{1}{c}{range (keV)} &  &  & \multicolumn{1}{c}{(\%)} \\
\hline
\hline
\noalign{\smallskip}
 7--30 & 5.5$\times$10$^{-13}$ & 4.1$\times$10$^{-10}$ & 0.13 \\
 15--150 & 1.5$\times$10$^{-13}$ & 3.8$\times$10$^{-10}$ & 0.039 \\
\noalign{\smallskip}
\hline
\hline
\end{tabular}
\end{center}
\end{table}

Following the approach in Baskill et al. (2005), we fit the
MECS spectral data using an absorbed optically thin thermal plasma
({\sc mekal} model in {\sc xspec}; Mewe et al. 1985). As no
spectral information was available below 2 keV in the MECS data,
we fixed the hydrogen column density to the value found by
Baskill et al. (2005), i.e. 2.8$\times$10$^{21}$ cm$^{-2}$.
The averaged spectrum was thus satisfactorily fitted ($\chi^2$/dof =
8.3/14) with a $kT$ = 8$^{+72}_{-4}$ (see Fig. A.2), in broad
agreement with the results of Baskill et al. (2005).
The best-fit model implied 7--30 keV and 15--150 keV fluxes for
the EY Cyg emission in the HPGSPC and PDS ranges as shown in Table A.1.

As one can see from this Table, the comparison between the fluxes
of 4U 1954+31 and EY Cyg indicates that any contamination from this 
field source on the spectrum of 4U 1954+31 is negligible.

\end{document}